\documentclass[twocolumn,showpacs,preprintnumbers,amsmath,amssymb]{revtex4}
\usepackage{amsmath}
\usepackage{graphicx}
\usepackage{dcolumn}
\usepackage{bm}

\newcommand{\ket}[1]{\mbox{$|#1\rangle$}}

\newcommand{\ketbra}[4]{\mbox{$|#1\rangle_{#2}\langle #3|_{#4}$}}

\begin{document}

\title{Proposal for optical parity state re-encoder}
\author{Y. X. Gong$^{1,2,}\protect\footnote{Electronic address: yxgong@mail.ustc.edu.cn}$, A. J. F. Hayes$^1$, G.
C. Guo$^2$ and T. C. Ralph$^1$\\
\emph{$^1$ Centre for Quantum Computer Technology, Department of Physics,}\\
\emph{University of Queensland, St Lucia 4072, Australia\\
 $^2$ Key
Laboratory of Quantum Information, University of Science
and Technology of China, CAS,\\
Hefei, 230026, People's Republic of China}}
\date{\today }

\begin{abstract}
We propose a re-encoder to generate a refreshed parity encoded state
from an existing parity encoded state. This is the simplest case of
the scheme by Gilchrist \emph{et al.} (Phys. Rev. A 75, 052328). We
show that it is possible to demonstrate with existing technology
parity encoded quantum gates and teleportation.
\end{abstract}

\pacs{03.67.Lx, 42.50.Dv, 42.79.Ta}
\maketitle

\section{Introduction}
Linear optical quantum computing with single photon modes (LOQC) was
novel and surprising, but not very practical when first introduced
some years ago \cite{KLM2001}. The idea was to restrict all in-line
processing to linear optical networks and to use measurement induced
non-linearities, additional single photon ancilla states and
feedforward to produce the necessary two-qubit interactions.
Although scalable in principle, the original scheme required an
unacceptably high number of operations ($\approx10,000$) to produce
a near deterministic two-qubit gate. More recently much progress has
been made in understanding and simplifying LOQC \cite{Kok2007}. In
particular two approaches, cluster states \cite{nielsen2004} and
parity states \cite{gilchrist2007}, can achieve near deterministic
two-qubit gates using approximately $100$ operations.

Although near deterministic operation is still some way off,
considerable success has been shown in demonstrating
non-deterministic optical quantum gates and small scale circuits
\cite{O'Brien2003, gasparoni2004, Pittman2003}. Progress has been
made in demonstrating the principles of cluster state computation
\cite{Walther2005, Prevedel2007}, but only basic parity state
demonstrations have so far been attempted \cite{o'brien2005,
pittman2005}.

Here we propose an experimental scheme that could demonstrate the
key features of parity state gate operation including loss error
detection \cite{ralph2005}, and that should be practical with
current down-conversion technology. The paper is arranged in the
following way. In the next section we introduce the parity encoding
and then describe the basic re-encoder demonstration in Section III.
In Section IV we consider gate operations and then follow in Section
V with a description of teleportation with the circuit. In Section
VI we talk about the possible experimental implementation using
parametric down-conversion processes, linear optical elements, and
conventional photon detectors. Section VII analyzes the effect that
optical mode-mismatch will have on the operation of the gates and we
conclude in Section VIII.

\section{Parity encoding}
Parity encoding has been shown as an efficient way to protect
against a computational basis measurement of one or more of the
component qubits \cite{KLM2001}. We will use the notation
$\ket{\Psi}^{(n)}$ to represent the logical state $\ket{\Psi}$ which
is parity encoded across $n$ distinct qubits. Explicitly, for an
arbitrary state $\ket{\Psi}=\alpha\ket{0}+\beta\ket{1}$, it can be
parity encoded as
$\ket{\Psi}^{(n)}=\alpha\ket{0}^{(n)}+\beta\ket{1}^{(n)}$. The even
and odd parity states are given by
\begin{eqnarray}
\ket{0}^{(n)}&\equiv&\frac{1}{\sqrt{2}}\left(\ket{+}^{\otimes n}+\ket{-}^{\otimes n}\right)\nonumber\\
\ket{1}^{(n)}&\equiv&\frac{1}{\sqrt{2}}\left(\ket{+}^{\otimes
n}-\ket{-}^{\otimes n}\right),
\end{eqnarray}
where$\ket{\pm}=(\ket{0}\pm\ket{1})/\sqrt{2}$. We can see that
$\ket{0}^{(n)}$ is represented as an equal superposition of all
states with even parity(the number of the component qubits in the
$\ket{1}$ state is even), and that $\ket{1}^{(n)}$ is represented as
an equal superposition of all states with odd parity(the number of
the component qubits in the $\ket{1}$ state is odd). If a
computational basis measurement is made on any of the component
qubits, it will not destroy the logical state, but collapse the
original state to
$\ket{\Psi}^{(n-1)}=\alpha\ket{0}^{(n-1)}+\beta\ket{1}^{(n-1)}$ if
the measurement result is ``$0$'', otherwise if the measurement
result is ``$1$'' the original state will collapse to
$\ket{\Phi}^{(n-1)}=\alpha\ket{1}^{(n-1)}+\beta\ket{0}^{(n-1)}$,
which can be corrected by a bit-flip on any of the remaining
component qubits.

In this paper, we consider the simplest parity encoded
state---two-qubit parity encoded state,
\begin{eqnarray}
\label{encoded}
\ket{\Psi}^{(2)}&=&\alpha\ket{0}^{(2)}+\beta\ket{1}^{(2)}\nonumber\\
&=&\frac{1}{\sqrt{2}}[\alpha(\ket{00}+\ket{11})+\beta(\ket{01}+\ket{10})].
\end{eqnarray}
We use the polarization states of a photon to construct the
component qubits, so that $\ket{0}\equiv\ket{H}$ and
$\ket{1}\equiv\ket{V}$.

A straightforward way \cite{Hayes2004} to prepare the parity encoded
state of an arbitrary state $\ket{\Psi}=\alpha\ket{H}+\beta\ket{V}$
is utilizing a controlled-not(CNOT) gate, with the photon in the
state $\ket{\Psi}$ as the target and an ancilla photon in the state
$(\ket{ H}+\ket{V})/\sqrt{2}$ as the control, which has been
experimentally demonstrated \cite{o'brien2005}. A simpler way to
generate a postselected parity encoded state is given by Pittman
\emph{et al.} \cite{pittman2005}, only using a single polarizing
beam splitter(PBS) and an ancilla photon. Here we give another way
to prepare the parity encoded state from the non-maximally entangled
state. To show this we rewrite Eq.~(\ref{encoded}) as
\begin{eqnarray}
\label{nonmaximal} \ket{\Psi}^{(2)}&=&\frac{1}{\sqrt{2}}[\alpha(\ket
{++}+\ket{
--})+\beta(\ket{++}-\ket{--})]\nonumber\\
&=&\frac{1}{\sqrt{2}}[(\alpha
+\beta)(\ket{++})+(\alpha-\beta)\ket{--}]\nonumber\\
&=&A\ket{++}+B\ket{--},
\end{eqnarray}
where $A\equiv(\alpha+\beta)/\sqrt{2}$,
$B\equiv(\alpha-\beta)/\sqrt{2}$, and $\ket{\pm}$ is the $\pm
45^\circ$ polarized state. Therefore we can first prepare a
non-maximally entangled state $A\ket{HH}+B\ket{VV}$,
 implement a Hadamard gate on each of the photons and then the parity
encoded state as shown in Eq.~(\ref{nonmaximal}) is yielded.

\section{The re-encoder}

An important technique in the application of parity encoding is the
re-encoder, i.e., to rebuild another parity encoded state from an
existing parity encoded state. Gilchrist \emph{et al.}
\cite{gilchrist2007} have given an efficient way using the
type-I($f_{\textrm{I}}$) fusion gate and the
type-II($f_{\textrm{II}}$) fusion gate,
\begin{eqnarray}
 Hf_{\textrm{I}}(H\otimes
H)\ket{0}^{(n)}\ket{0}^{(m)}\rightarrow\left\{
\begin{array}{cl}\ket {0}^{(m+n-1)} &\mbox {(success)}\\ - &\mbox
{(failure)}\end{array}\right.\nonumber\\
{\label{type-I}}\\
 f_{\textrm{II}}\ket{0}^{(n)}\ket{0}^{(m)}\rightarrow\left\{
\begin{array}{cl}\ket{0}^{(m+n-2)} &\mbox{(success)}\\
\ket{0}^{(m-1)}\ket{0}^{(n-1)} &\mbox
{(failure)}\end{array},\right.\nonumber\\
{\label{type-II}}
\end{eqnarray}
and each fusion succeeds with a probability of $1/2$. In this paper
we give a detailed analysis of the simplest case and investigate the
property of the re-encoder.

\begin{figure}[htb]
\centering
\includegraphics[width=9.2cm]{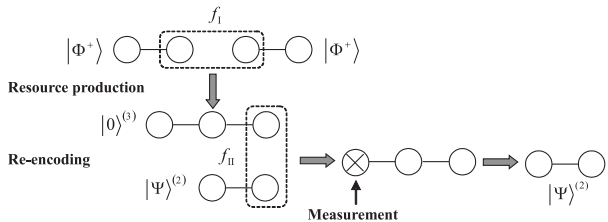}
\caption{Schematic of the re-encoder.} \label{fig_schematic}
\end{figure}

\begin{figure}[htb]
\centering
\includegraphics[width=8cm]{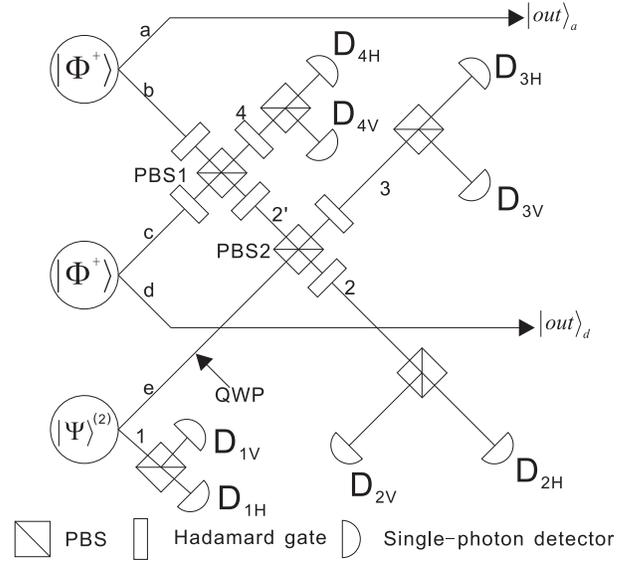}
\caption{Scheme to implement the two-photon parity state re-encoder.
Lowercase letters and numbers label the beams. Polarizing beam
splitters transmit horizontally polarized photons$(\ket{H})$ and
reflect vertically polarized photons$(\ket{V})$. Hadamard gates
transform horizontally polarized photons to $+45^{\circ}$-polarized
photons $\ket{+}$ and transform vertically polarized photons to
$-45^{\circ}$-polarized photons$(\ket{-})$, which can be implemented
by a half-wave plate oriented at $22.5^{\circ}$. A quarter-wave
plate(QWP) set to $0^{\circ}$ is inserted in beam $e$ for
implementing a $Z_{90}$ gate on the photon in mode $e$. }
\label{fig_Experiment}
\end{figure}

Fig.\ref{fig_schematic} shows the schematic of the re-encoder. We
start from three entangled-photon pairs, two of which are in the
Bell state \mbox{$\ket{\Phi^{+}}=(\ket{HH}+\ket{VV})\sqrt{2}$}, with
the third one in the parity encoded state as shown in
Eq.~(\ref{nonmaximal}). We first assume that we have ideal
entangled-photon sources and in Section VI we will talk about the
generation of the entangled-photon pairs in real experiment. First
we generate the resource state $\ket{0}^{(3)}$ from two Bell states
using the $f_{\textrm{I}}$ fusion gate. Second we fuse the original
parity encoded state and the resource state using the
$f_{\textrm{II}}$ fusion gate. Third after projection measurement on
the remaining photon of the original parity encoded state we produce
a new parity encoded state.

Fig.\ref{fig_Experiment} shows the experimental scheme to
demonstrate the re-encoder. To show the operation of the re-encoder
we first consider the fusion between the two Bell states
$\ket{\Phi^{+}}_{ab}$ and $\ket{\Phi^{+}}_{cd}$ at PBS1. The state
transformation before PBS2 can be written as
\begin{eqnarray}
\label{PBS1} \ket{\Phi^{+}}_{ab}\ket{\Phi^{+}}_{cd}\rightarrow\hspace{3cm}\nonumber\\
\frac{1}{4}\left[\ket{H}_{4}\left(\ket{H}_{a}\ket{H}_{d}\ket{H}_{2'}\right.\right.&+&\ket{H}_{a}\ket{V}_{d}\ket{V}_{2'}\nonumber\\
+\ket{V}_{a}\ket{V}_{d}\ket{H}_{2'}&+&\left.\ket{V}_{a}\ket{H}_{d}\ket{V}_{2'}\right)\nonumber\\
+\ket{V}_{4}\left(\ket{V}_{a}\ket{H}_{d}\ket{H}_{2'}\right.&+&\ket{H}_{a}\ket{V}_{d}\ket{H}_{2'}\nonumber\\
+\ket{H}_{a}\ket{H}_{d}\ket{V}_{2'}&+&\left.\left.\ket{V}_{a}\ket{V}_{d}\ket{V}_{2'}\right)\right]\nonumber\\
+\frac{1}{\sqrt{2}}\ket{\varphi_{\textrm{I}}},\hspace{2cm}
\end{eqnarray}
where $\ket{\varphi_{\textrm{I}}}$ is a normalized combination of
all the amplitudes that would not lead to exactly one photon in
detectors $D_{4H}$ and $D_{4V}$. If the detector $D_{4H}$ receives
exactly one photon the state in mode $a$, $d$, and $2'$ will
collapse to $\ket{0}^{(3)}$, while if there is exactly one photon in
$D_{4V}$ the state
 will be $\ket{1}^{(3)}$, with a probability of $1/4$ for each
 result.

After the state $\ket{0}^{(3)}(\ket{1}^{(3)})$ is yielded, the
photons in beams $2'$ and $e$ are sent to PBS2, followed by
detectors in beams $2$ and $3$, which act as a type-II fusion gate
together with the PBS2. The whole state transformation can be
written as
\begin{eqnarray}
\label{PBS2} \ket{\Phi^{+}}_{ab}\ket{\Phi^{+}}_{cd}\ket{\Psi}^{(2)}_{e1}\rightarrow\hspace{5cm}\nonumber\\
\frac{1}{8\sqrt{2}}\left\{\left(\ket{H}_{1}\ket{H}_{2}\ket{H}_{3}\ket{H}_{4}\right.\right.+\ket{H}_{1}\ket{V}_{2}\ket{V}_{3}\ket{H}_{4}\nonumber\hspace{0.5cm}\\
+\ket{V}_{1}\ket{H}_{2}\ket{H}_{3}\ket{V}_{4}+\left.\ket{V}_{1}\ket{V}_{2}\ket{V}_{3}\ket{V}_{4}\right)\nonumber\hspace{0.35cm}\\
\otimes\left[\alpha\left(\ket{H}_{a}\ket{H}_{d}+\ket{V}_{a}\ket{V}_{d}\right)\right.\hspace{2.5cm}\nonumber\\
+\beta\left.\left(\ket{H}_{a}\ket{V}_{d}+\ket{V}_{a}\ket{H}_{d}\right)\right]\hspace{2.2cm}\nonumber\\
+\left(\ket{H}_{1}\ket{V}_{2}\ket{H}_{3}\ket{H}_{4}\right.+\ket{H}_{1}\ket{H}_{2}\ket{V}_{3}\ket{H}_{4}\nonumber\hspace{0.5cm}\\
-\ket{V}_{1}\ket{V}_{2}\ket{H}_{3}\ket{V}_{4}-\left.\ket{V}_{1}\ket{H}_{2}\ket{V}_{3}\ket{V}_{4}\right)\nonumber\hspace{0.35cm}\\
\otimes\left[\alpha\left(\ket{H}_{a}\ket{H}_{d}+\ket{V}_{a}\ket{V}_{d}\right)\right.\hspace{2.5cm}\nonumber\\
-\beta\left.\left(\ket{H}_{a}\ket{V}_{d}+\ket{V}_{a}\ket{H}_{d}\right)\right]\hspace{2.2cm}\nonumber\\
+\left(\ket{V}_{1}\ket{H}_{2}\ket{H}_{3}\ket{H}_{4}\right.+\ket{V}_{1}\ket{V}_{2}\ket{V}_{3}\ket{H}_{4}\nonumber\hspace{0.5cm}\\
+\ket{H}_{1}\ket{H}_{2}\ket{H}_{3}\ket{V}_{4}+\left.\ket{H}_{1}\ket{V}_{2}\ket{V}_{3}\ket{V}_{4}\right)\nonumber\hspace{0.35cm}\\
\otimes\left[\alpha\left(\ket{H}_{a}\ket{V}_{d}+\ket{V}_{a}\ket{H}_{d}\right)\right.\hspace{2.5cm}\nonumber\\
+\beta\left.\left(\ket{H}_{a}\ket{H}_{d}+\ket{V}_{a}\ket{V}_{d}\right)\right]\hspace{2.2cm}\nonumber\\
+\left(\ket{H}_{1}\ket{V}_{2}\ket{H}_{3}\ket{V}_{4}\right.+\ket{H}_{1}\ket{H}_{2}\ket{V}_{3}\ket{V}_{4}\nonumber\hspace{0.5cm}\\
-\ket{V}_{1}\ket{V}_{2}\ket{H}_{3}\ket{H}_{4}-\left.\ket{V}_{1}\ket{H}_{2}\ket{V}_{3}\ket{H}_{4}\right)\nonumber\hspace{0.35cm}\\
\otimes\left[\alpha\left(\ket{H}_{a}\ket{V}_{d}+\ket{V}_{a}\ket{H}_{d}\right)\right.\hspace{2.5cm}\nonumber\\
-\beta\left.\left.\left(\ket{H}_{a}\ket{H}_{d}+\ket{V}_{a}\ket{V}_{d}\right)\right]\right\}\hspace{2.2cm}\nonumber\\
+\frac{\sqrt{3}}{2}\ket{\varphi_{\textrm{II}}},\hspace{5.5cm}
\end{eqnarray}
where $\ket{\varphi_{\textrm{II}}}$ is a normalized state including
the amplitudes that would lead to unsuccessful cases in which one or
more of the eight detectors detect more than one photon.

From Eq.~{(\ref{PBS2})} it is clear that a different combination of
the detectors receiving exactly one photon leads to a different
output state in modes $a$ and $d$ as follows,
\begin{eqnarray}
\label{output1}
&&\ket{\Psi_1}_{ad}=\alpha\ket{0}^{(2)}+\beta\ket{1}^{(2)},\\
\label{output2}
&&\ket{\Psi_2}_{ad}=\alpha\ket{0}^{(2)}-\beta\ket{1}^{(2)},\\
\label{output3}
&&\ket{\Psi_3}_{ad}=\alpha\ket{1}^{(2)}+\beta\ket{0}^{(2)},\\
\label{output4}
&&\ket{\Psi_4}_{ad}=\alpha\ket{1}^{(2)}-\beta\ket{0}^{(2)}.
\end{eqnarray}

For example, combination of $D_{1H}$, $D_{2H}$, $D_{3H}$ and
$D_{4H}$ gives the state $\ket{\Psi_1}_{ad}$ as shown in
Eq.~(\ref{output1}), while combination of $D_{1V}$, $D_{2V}$,
$D_{3H}$ and $D_{4H}$ gives the state $\ket{\Psi_4}_{ad}$ as shown
in Eq.~(\ref{output4}). Although the four output states are
different, they are locally equivalent. The state
$\ket{\Psi_1}_{ad}$ is the expected outcome, i.e. equivalent to
$\ket{\Psi}^{(2)}_{e1}$. The state $\ket{\Psi_2}_{ad}$ is only a
phase-flipped version of $\ket{\Psi_1}_{ad}$, which can be
transformed to $\ket{\Psi_1}_{ad}$ by implementing the local
operation $\sigma_z$ on both the photons in modes $a$ and $d$. For
the state $\ket{\Psi_3}_{ad}$, a bit-flip occurs, which can be
corrected by a local operation $\sigma_x$ on either of the photons
in modes $a$ and $d$. Both a bit-flip and a phase-flip occur to the
state $\ket{\Psi_4}_{ad}$, so the local operation is
$\sigma_{z_a}\otimes\sigma_{z_d}\sigma_{x_d}$ or
$\sigma_{z_a}\sigma_{x_a}\otimes\sigma_{z_d}$.

Eq.~{(\ref{PBS2})} also shows that each output state succeeds with a
probability of $1/16$, however, if we accept all the four states and
use the classically controlled single-qubit operations as we have
shown, then the probability can be increased to $1/4$. In a
scaled-up scheme fusion with the qubit state would only proceed when
the resource had been successfully constructed. In addition, because
of the characteristics of the type-II fusion gate (See
Eq.~{(\ref{type-II})}), recovery from unsuccessful fusion of the
parity qubit is also possible. Iteration can then improve the
probability of success towards unity. Such scale-up would require
efficient multi-photon production and quantum memory beyond current
capabilities, and hence here we restrict our attention mostly to the
basic probabilistic operations (see Section V for a discussion of
recovery techniques in a teleportation scenario).

\section{Parity encoded quantum computation}
It is known that with polarization encoded qubits we can perform any
single-qubit unitary operation deterministically with passive linear
optical elements. Gates between different photons, like the CNOT
gate, can be performed nondeterministically. However, performing
gates on the parity encoded states is somewhat different.

It is straightforward to perform any of the gates which can be
achieved with the set $\{X_\theta, Z\}$ on the parity encoded
states. Here the notation is
$X_\theta=\cos(\theta/2)I-i\sin(\theta/2)\sigma_x$ and $Z$ means the
$\sigma_z$ operation. To perform an $X_\theta$ rotation on a parity
encoded state, we only need to perform the rotation on any of the
component qubits because the $\sigma_x$ rotation on any of the
component qubits can change the parity of the encoded state. To
perform a $Z$ operation, we need to perform the $\sigma_z$ operation
on all the component qubits as the odd parity states will suffer an
overall phase-flip. However, to achieve a universal set of gates we
need to add the set $\{Z_{90},CNOT\}$. Here the notation is
$Z_{90}=e^{-i\pi\sigma_{z}/4}$. We can only perform these gates on
parity encoded states nondeterministically.

According to the proposal by Gilchrist \emph{et al.}
\cite{gilchrist2007}, we give a detailed analysis of how to
implement the $Z_{90}$ gate with the re-encoder. The main procedure
is that we first perform the $Z_{90}$ gate on one of the component
qubits and then re-encode from that qubit. By inserting a
quarter-wave plate (QWP) set to $0^{\circ}$ in beam $e$ prior to
PBS2, we can perform the $Z_{90}$ gate on the photon in mode $e$.
Then the state in modes $e$ and $1$ will be transformed to
\begin{eqnarray}
\label{Z90}
\ket{\Psi'}_{e1}=\frac{1}{\sqrt{2}}[\alpha(\ket{H}_e\ket{H}_1&+&i\ket{V}_e\ket{V}_1)\nonumber\\
+\beta(\ket{H}_e\ket{V}_1&+&i\ket{V}_e\ket{H}_1)].
\end{eqnarray}
Based on similar calculations as those used in the description of
the re-encoder, we can also get four different output states in
modes $a$ and $d$ corresponding to different combinations of the
detectors receiving exactly one photon, but the corrections are
different for some of the combinations.

For the combination $D_{1H}$,$D_{2H}$,$D_{3H}$,$D_{4H}$, or
$D_{1H}$,$D_{2V}$,$D_{3V}$,$D_{4H}$, or
$D_{1V}$,$D_{2V}$,$D_{3H}$,$D_{4V}$, or
$D_{1V}$,$D_{2H}$,$D_{3V}$,$D_{4V}$, the output state collapses to
\begin{eqnarray}
\label{Z90:1}
\ket{\Psi_1'}_{ad}=\alpha\ket{0}^{(2)}+i\beta\ket{1}^{(2)},
\end{eqnarray}
which is the expected state. If the combination is
$D_{1V}$,$D_{2H}$,$D_{3H}$,$D_{4V}$, or
$D_{1V}$,$D_{2V}$,$D_{3V}$,$D_{4V}$, or
$D_{1H}$,$D_{2V}$,$D_{3H}$,$D_{4H}$, or
$D_{1H}$,$D_{2H}$,$D_{3V}$,$D_{4H}$, the output state will become to
\begin{eqnarray}
\label{Z90:2}
\ket{\Psi_2'}_{ad}=\alpha\ket{0}^{(2)}-i\beta\ket{1}^{(2)},
\end{eqnarray}
which is a phase-flipped version, so we need a local operation
$\sigma_z$ on both of the two output modes. The combination
$D_{1H}$,$D_{2H}$,$D_{3H}$,$D_{4V}$, or
$D_{1H}$,$D_{2V}$,$D_{3V}$,$D_{4V}$, or
$D_{1V}$,$D_{2V}$,$D_{3H}$,$D_{4H}$, or
$D_{1V}$,$D_{2H}$,$D_{3V}$,$D_{4H}$ leads to the output state
\begin{eqnarray} \label{Z90:3}
\ket{\Psi_3'}_{ad}=\alpha\ket{1}^{(2)}+i\beta\ket{0}^{(2)},
\end{eqnarray}
which only needs a correction of a local operation $\sigma_x$ on
either of the two output modes. Another combination is
$D_{1V}$,$D_{2H}$,$D_{3H}$,$D_{4H}$, or
$D_{1V}$,$D_{2V}$,$D_{3V}$,$D_{4H}$, or
$D_{1H}$,$D_{2V}$,$D_{3H}$,$D_{4V}$, or
$D_{1H}$,$D_{2H}$,$D_{3V}$,$D_{4V}$, which results in the output
state
\begin{eqnarray} \label{Z90:4}
\ket{\Psi_4'}_{ad}=\alpha\ket{1}^{(2)}-i\beta\ket{0}^{(2)},
\end{eqnarray}
therefore we need a correction
$\sigma_{z_a}\otimes\sigma_{z_d}\sigma_{x_d}$ or
$\sigma_{z_a}\sigma_{x_a}\otimes\sigma_{z_d}$. The probability of
each success output state is again $1/16$, which can also be
improved to $1/4$ if we accept all the four output states.

For the operation of the CNOT gate, Gilchrist et al.
\cite{gilchrist2007} proposed a procedure similar to the $Z_{90}$
gate. We will not show the detail of the CNOT implementation:
\begin{eqnarray}
\ket{\Psi}_c^{(n)}\ket{\Psi}_t^{(n)}\rightarrow \mbox{CNOT}
\ket{\Psi}_c^{(n)}\ket{\Psi}_t^{(n)},
\end{eqnarray}
where $\ket{\Psi}_c^{(n)}(\ket{\Psi}_t^{(n)})$ is the $n$-qubit
parity encoded control (target) state. Again using a type-I fusion
gate and a type-II fusion gate we can first implement the operation:
\begin{eqnarray}
\ket{\Psi}_c^{(n)}\ket{\Psi}_t^{(n)}\ket{0}^{(n+1)}\rightarrow
\mbox{CNOT}\ket{\Psi}_c^{(n)}\ket{\Psi}_t^{(n-1)},
\end{eqnarray}
and then a re-encoder can encode the target state to a $n$-qubit
parity encoded state. However note that for the first level parity
state ($n=2$), far more than six photons are needed and hence it is
beyond the scope of this paper.

\section{Parity encoded quantum teleportation}
Another way to understand the re-encoder is in terms of
teleportation. Quantum teleportation \cite{Bennett1993}, a way to
transfer a quantum state from one place to another, has received
much attention since it was presented. It plays an important role in
quantum communication \cite{Briegel1998} and computation
\cite{Gottesman1999,KLM2001}. An efficient way to improve the
probability of success of teleportation is using parity encoding to
encode against the failure which results in a computational basis
measurement \cite{KLM2001,Kok2007}. Quantum teleportation of an
arbitrary two-qubit composite system has been realized in the
experiment \cite{Zhang2006}. This experiment demonstrated
teleportation of a two-qubit parity encoded state, but it couldn't
show all the features against failure because the Bell measurement
was implemented on each of the two qubits individually. Here we give
another way to teleport a parity encoded state using the re-encoder,
which demonstrates the basic ability to encode against the failure
in teleportation.

As illustrated in Fig.\ref{fig_Experiment}, Alice wants to teleport
an unknown parity encoded state $\ket{\Psi}^{(2)}_{e1}$ in modes $e$
and $1$ to Bob. To do so, first Alice and Bob share two Bell states
$\ket{\Phi^+}_{ab}$ and $\ket{\Phi^+}_{cd}$, where the photons in
modes $b$ and $c$ are sent to Alice while the photons in modes $a$
and $d$ are sent to Bob. Alice then carries out the fusion
operations used in the re-encoder and tells Bob the measurement
results in modes $1$, $2$, $3$ and $4$ via classical communication.
On receiving these results, with the corrections shown in the
description of the re-encoder, Bob can then get the state in modes
$a$ and $d$, which is the same with the teleported state by Alice.
As in the re-encoder, the total probability of success of
teleportation is also $1/4$.

To explore the procedure of encoding against the failure, we analyze
the failure cases in the two fusion gates. If a failure occurs to
the measurement in mode $4$, Alice can try again the fusion at PBS1
instead of carrying out the fusion at PBS2, so that the state in
modes $e$ and $1$ will not be destroyed and can still be used unless
the measurement in mode $4$ succeeds. If a failure occurs to the
fusion at PBS2, it will destroy the state in modes $e$ and $1$, but
that is not a problem for Alice, because she need not make a
measurement in mode $1$. Instead, she need only re-encode from the
state in mode $1$, with only a bit-flip correction required
depending on the failure results. Note that the proposal in this
paper can't be used to recover the initial state from mode $1$,
because we do not use an encoder to generate the parity encoded
state, but that does not affect the role of the re-encoder.

The procedure of recycling of entangled states against the failure
can also be implemented for the parity encoded quantum computation,
as was done for the cluster computation proposal \cite{Browne2005}.

\section{Generation of the three entangled states using parametric down-conversion processes}
So far we have considered the situation in which our source
deterministically produces three entangled-photon Bell pairs.
However in real experiment it may not be the case. Currently, nearly
all the entangled-photon sources in LOQC experiments use parametric
down-conversion(PDC) \cite{Kwiat1995,Kwiat1999}. In order to
demonstrate the re-encoder we need three entangled-photon pairs, so
a six-photon source is required, which is now available in the
experiment \cite{Zhang2006,lu2007}. We may use three PDC sources,
each of which generates an entangled-photon pair.

As we know, PDC is a multi-photon generation process. The
probabilities that a $n$-pair is generated from a single PDC source
is the same as that a single pair is generated from $n$ PDC sources
\cite{Walther2005}. If we first consider three-pair order, we see
that when a double-pair is produced in one of the Bell state sources
while no pairs are produced in another we can also get the correct
fourfold detections but the output states are wrong results with two
photons in one of the output modes while no photons in another.
However, by specifically using the optical arrangement of
Fig.\ref{fig_Experiment}, and using post-selection, i.e., sixfold
coincidence detections of the correct fourfold detections together
with the detections of the two output modes to make sure a photon
exits in each output mode, can solve this problem. This type of
method is currently used in virtually all experiments using PDC.

Another concern is due to higher-order processes in which more than
six photons are generated and this can also lead to wrong results
even though we use post-selection. In real experiment, the
efficiency of generating two photons per pulse from PDC is typically
$\vert\chi\vert^2\sim10^{-4}$. Then the efficiency of six- and
eight-photon generation from PDC is $\vert\chi\vert^6\sim10^{-12}$
and $\vert\chi\vert^8\sim10^{-16}$. Therefore the eight-photon
generation rate is $\sim10^{-4}$ lower than that of six-photon
generation and is negligible. It should be noted that theoretically
we assume the detectors are number-resolving, however, in the case
of post-selection, conventional detectors are acceptable, since
higher-order processes are negligible. Another error source in a
real experiment is dark counts of conventional detectors, but that
of current detectors is sufficiently small and hence they are
negligible in multi-photon coincidence experiments.

From all the discussions above, we believe that with existing
technology the proposed re-encoder is able to be performed using PDC
sources, linear optical elements, and conventional photon detectors.

\section{Mode-mismatch errors}
In nonclassical interference experiments a major contribution of
nonunit visibility is mode-mismatch. To model mode-mismatch explicit
multi-mode calculations may be used \cite{Rohde2005a}, but it
becomes very complicated to deal with multi-photon set-ups. In this
paper we model mode-mismatch using a simpler approach similar to the
analysis by Ralph \emph{et al.} \cite{Ralph2002}. A rigorous
justification of this approach can be found in \cite{Rohde2007}.

As shown in Fig.\ref{fig_Experiment}, there is nonclassical
interference at PBS1 and PBS2. We introduce two parameters $\eta_1$
and $\eta_2$ $(0\leqslant\eta_1,\eta_2\leqslant1)$ to quantify the
degrees of mode matching between the two input modes of PBS1 and
PBS2. We assume that due to mode-mismatch the three initial
entangled states become {\setlength\arraycolsep{2pt}
\begin{eqnarray}\label{mismatch}
&\ket{\Phi^+}_{ab}&\ket{\Phi^+}_{cd}\ket{\Psi}^{(2)}_{e1}\rightarrow\nonumber\\
&&\left(\sqrt{\eta_1}\ket{\Phi^+}_{ab}+\sqrt{1-\eta_1}\ket{\Phi^+}'_{ab}\right)\ket{\Phi^+}_{cd}\nonumber\\
&&\otimes\left(\sqrt{\eta_2}\ket{\Psi}^{(2)}_{e1}+\sqrt{1-\eta_2}\ket{\Psi}{''}^{(2)}_{e1}\right)\nonumber\\
&=&\sqrt{\eta_1\eta_2}\ket{\Phi^+}_{ab}\ket{\Phi^+}_{cd}\ket{\Psi}^{(2)}_{e1}\nonumber\\
&&+\sqrt{(1-\eta_1)(1-\eta_2)}\ket{\Phi^+}'_{ab}\ket{\Phi^+}_{cd}\ket{\Psi}{''}^{(2)}_{e1}\nonumber\\
&&+\sqrt{\eta_1(1-\eta_2)}\ket{\Phi^+}_{ab}\ket{\Phi^+}_{cd}\ket{\Psi}{''}^{(2)}_{e1}\nonumber\\
&&+\sqrt{(1-\eta_1)\eta_2}\ket{\Phi^+}'_{ab}\ket{\Phi^+}_{cd}\ket{\Psi}^{(2)}_{e1}.
\end{eqnarray}}
Here
\begin{eqnarray}
\label{mismatchab}
\ket{\Phi^+}'_{ab}&=&\frac{1}{\sqrt{2}}\left(\ket{H}_a\ket{H}'_b+\ket{V}_a\ket{V}'_b\right),\\
\label{mismatche1}
\ket{\Psi}'^{(2)}_{e1}&=&\frac{1}{\sqrt{2}}\left[\alpha\left(\ket{H}'_{e}\ket{H}_1+\ket{V}'_{e}\ket{V}_1\right)\right.\nonumber\\
&&\qquad+\left.\beta\left(\ket{H}'_{e}\ket{V}_1+\ket{V}'_{e}\ket{H}_1\right)\right],
\end{eqnarray}
where we use the notation $\ket{H}'_b\left(\ket{V}'_b\right)$ to
denote the mode-mismatch state of the photon in mode $b$, which is
distinguishable from the state $\ket{H}_c\left(\ket{V}_c\right)$ of
the photon in mode $c$. Similarly the mode-mismatch state
$\ket{H}{''}_e\left(\ket{V}{''}_e\right)$ is distinguishable from
$\ket{H}_d\left(\ket{V}_d\right)$. Note that although we only
introduce the mode-mismatch in a single degree of freedom, this is
sufficient to model arbitrary mode-mismatch effects
\cite{Rohde2005b}.

We can see that there are four distinguishable cases corresponding
to the four terms of Eq.~(\ref{mismatch}) and the state
transformation for each of them can be given in the same way used in
the description of the re-encoder. For simplicity we only give the
states which can lead to exactly one photon in the detectors
$D_{1H}$, $D_{2H}$, $D_{3H}$ and $D_{4H}$, so the four terms of
Eq.~(\ref{mismatch}) evolve such that,\\ the first term:
\begin{eqnarray}\label{mismatch1}
\sqrt{\eta_1\eta_2}\ket{\Phi^+}_{ab}\ket{\Phi^+}_{cd}\ket{\Psi}^{(2)}_{e1}\rightarrow\hspace{3.5cm}\nonumber\\
\frac{1}{8}\sqrt{\eta_1\eta_2}\ket{H}_1\ket{H}_2\ket{H}_3\ket{H}_4\ket{\Psi}^{(2)}_{ad}\:,\hspace{0.5cm}
\end{eqnarray}
the second term:
\begin{eqnarray*}
\sqrt{(1-\eta_1)(1-\eta_2)}\ket{\Phi^+}'_{ab}\ket{\Phi^+}_{cd}\ket{\Psi}{''}^{(2)}_{e1}\rightarrow\hspace{4.5cm}\nonumber\\
\frac{1}{8\sqrt{2}}\sqrt{(1-\eta_1)(1-\eta_2)}\left[\alpha\ket{H}_1\ket{H}'_2\ket{H}{''}_3\ket{H}_4\ket{+}_a\ket{+}_d\right.\hspace{3cm}\nonumber\\
+\beta\ket{H}_1\ket{H}{''}_2\ket{H}{'}_3\ket{H}_4\ket{+}_a\ket{+}_d\hspace{3cm}\nonumber\\
+\alpha\ket{H}_1\ket{H}_2\ket{H}{''}_3\ket{H}'_4\ket{-}_a\ket{-}_d\hspace{3.1cm}\nonumber\\
\left.-\beta\ket{H}_1\ket{H}{''}_2\ket{H}_3\ket{H}'_4\ket{-}_a\ket{-}_d\right],\hspace{2.7cm}\nonumber\\
\end{eqnarray*}
\begin{equation}
\label{mismatch2}
\end{equation}
the third term:
\begin{eqnarray}\label{mismatch3}
\sqrt{\eta_1(1-\eta_2)}\ket{\Phi^+}_{ab}\ket{\Phi^+}_{cd}\ket{\Psi}{''}^{(2)}_{e1}\rightarrow\hspace{2cm}\nonumber\\
\frac{1}{8}\sqrt{\eta_1(1-\eta_2)}\left[\alpha\ket{H}_1\ket{H}_2\ket{H}{''}_3\ket{H}_4\ket{0}^{(2)}_{ad}\right.\hspace{0.5cm}\nonumber\\
\left.+\beta\ket{H}_1\ket{H}{''}_2\ket{H}_3\ket{H}_4\ket{1}^{(2)}_{ad}\right],\hspace{0.3cm}
\end{eqnarray}
the fourth term:
\begin{eqnarray} \label{mismatch4}
\sqrt{(1-\eta_1)\eta_2}\ket{\Phi^+}'_{ab}\ket{\Phi^+}_{cd}\ket{\Psi}^{(2)}_{e1}\rightarrow\hspace{2.6cm}\nonumber\\
\frac{1}{8\sqrt{2}}\sqrt{(1-\eta_1)\eta_2}\left[\alpha\ket{H}_1\ket{H}'_2\ket{H}_3\ket{H}_4\ket{+}_a\ket{+}_d\right.\hspace{1.431cm}\nonumber\\
+\beta\ket{H}_1\ket{H}_2\ket{H}'_3\ket{H}_4\ket{+}_a\ket{+}_d\hspace{1.43cm}\nonumber\\
\left.+(\alpha-\beta)\ket{H}_1\ket{H}_2\ket{H}_3\ket{H}'_4\ket{-}_a\ket{-}_d\right],\hspace{0.2cm}
\end{eqnarray}
from which we can see that the first term leads the output state in
modes $a$ and $d$ to the expected state $\ket{\Psi}^{(2)}_{ad}$,
while the other terms lead to a mixed state in modes $a$ and $d$.
For other combinations of detectors as shown in the analysis of the
re-encoder we can also get similar results with corresponding
corrections.

Based on these observations, the density matrix of the output state
can be given by
\begin{eqnarray}
\hat{\rho}^{(\pm)}_{out}=\frac{1}{64}\eta_1\eta_2\ket{\Psi}^{(2)}_{ad}\langle\Psi\vert^{(2)}_{ad}\hspace{4cm}\nonumber\\
+\frac{1}{128}(1-\eta_1)\ketbra{\pm}{a}{\pm}{a}\otimes\ketbra{\pm}{d}{\pm}{d}\hspace{1.6cm}\nonumber\\
+\frac{1}{128}(1-\eta_1)\left[1\mp2\textrm{Re}\left(\alpha\beta^*\right)\eta_2\right]\hspace{1.9cm}\nonumber\\
\times\ketbra{\mp}{a}{\mp}{a}\otimes\ketbra{\mp}{d}{\mp}{d}\hspace{1.5cm}\nonumber\\
+\frac{1}{64}\eta_1(1-\eta_2)\vert\alpha\vert^2\ket{0}^{(2)}_{ad}\langle0\vert^{(2)}_{ad}\hspace{2.45cm}\nonumber\\
+\frac{1}{64}\eta_1(1-\eta_2)\vert\beta\vert^2\ket{1}^{(2)}_{ad}\langle1\vert^{(2)}_{ad},\hspace{2.35cm}
\end{eqnarray}
where $\hat{\rho}^{(+)}_{out}$ corresponds to the output state with
no correction or with the bit-flip correction, while
$\hat{\rho}^{(-)}_{out}$ corresponds to the output state after the
phase-flip correction or after both the phase-flip and the bit-flip
correction.

The probability of success is given by
\begin{eqnarray}\label{Probability}
P^{(\pm)}=\mbox{tr}\left(\hat{\rho}^{(\pm)}_{out}\right)=\frac{1}{64}[1{\mp}(1-\eta_1)\eta_2\mbox{Re}(\alpha\beta^*)].
\end{eqnarray}

The fidelity is given by
\begin{eqnarray}\label{fidelity}
F^{(\pm)}=\frac{\langle\Psi\vert^{(2)}\hat{\rho}^{(\pm)}_{out}\ket{\Psi}^{(2)}}{\mbox{tr}\left(\hat{\rho}^{(\pm)}_{out}\right)},
\end{eqnarray}
where,
\begin{eqnarray}\label{fid}
\langle\Psi\vert^{(2)}\hat{\rho}^{(\pm)}_{out}\ket{\Psi}^{(2)}=\hspace{5cm}\nonumber\\
\frac{1}{128}\left\{1+\eta_1-4\eta_1(1-\eta_2)\vert\alpha\beta\vert^2\right.\hspace{2cm}\nonumber\\
\left.{\mp}(1-\eta_1)\eta_2\mbox{Re}\left(\alpha\beta^*\right)\left[1{\mp}2\mbox{Re}\left(\alpha\beta^*\right)\right]\right\}.\
\end{eqnarray}

From Eqs.~(\ref{Probability}), (\ref{fidelity}) and (\ref{fid}) we
can see that the fidelity depends on not only the mode matching
parameters but also the input states. If we integrate over the whole
space of the pure input state, we find that the average fidelity is
the same for $F^{(\pm)}$. To show that, we make the substitution,
\begin{eqnarray}
\label{alpha} \alpha&=&\cos\frac{\theta}{2},\\
 \label{beta}
\beta&=&e^{i\phi}\sin\frac{\theta}{2}.
\end{eqnarray}
The average fidelity can then be given by
\begin{eqnarray}\label{average}
F_{\textrm{ave}}&=&\frac{1}{4\pi}\int^{2\pi}_{0}d\phi\int^{\pi}_{0}F^{(\pm)}\sin{\theta}d\theta\nonumber\\
&=&\frac{1}{4\pi}\int^{\frac{\pi}{2}}_{-\frac{\pi}{2}}d\phi\int^{\pi}_{0}\left(F^{(+)}+F^{(-)}\right)\sin{\theta}d\theta.
\end{eqnarray}

\begin{figure}[htb]
%\begin{center}
\centering
\includegraphics[width=8cm]{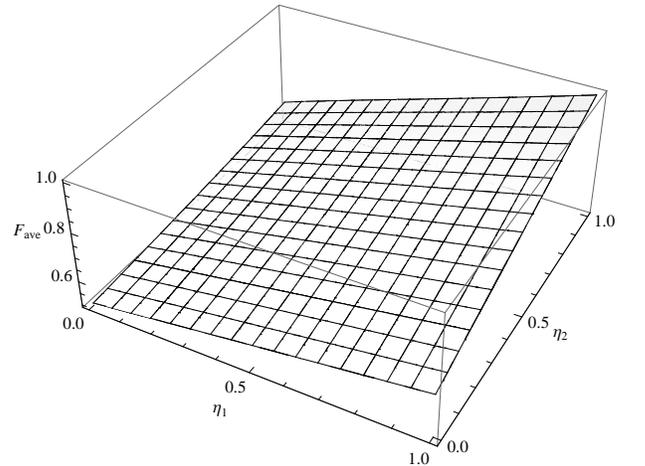}
\caption{Average fidelity $F_{\textrm{ave}}$ as a function of mode
matching parameters $\eta_1$ and $\eta_2$.} \label{fig_fidelity3D}
%\end{center}
\end{figure}

\begin{figure}[htb]
%\begin{center}
\centering
\includegraphics[width=8cm]{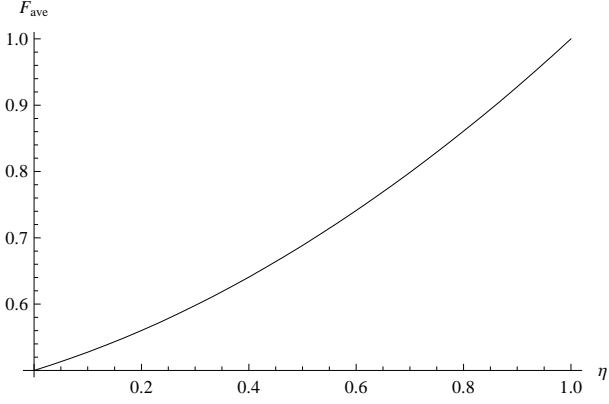}
\caption{Average fidelity $F_{\textrm{ave}}$ as a function of
$\eta$, assuming mode matching parameters $\eta_1=\eta_2=\eta$.}
\label{fig_fidelity2D}
%\end{center}
\end{figure}

In Fig.\ref{fig_fidelity3D} we plotted the average fidelity as a
function of mode matching parameters $\eta_1$ and $\eta_2$. If we
assume \mbox{$\eta_1=\eta_2=\eta$}, the relationship between the
average fidelity and $\eta$ is shown in Fig.\ref{fig_fidelity2D}.

Similar analysis of mode-mismatch can also be implemented to the
$Z_{90}$ operation. The density matrix of the output state after the
$Z_{90}$ operation is given by
\begin{eqnarray}
\widetilde{\hat{\rho}}^{(\pm)}_{out}=\frac{1}{64}\eta_1\eta_2\ket{\Psi_1'}_{ad}\langle\Psi_1'\vert_{ad}\hspace{4cm}\nonumber\\
+\frac{1}{128}(1-\eta_1)\ketbra{\pm}{a}{\pm}{a}\otimes\ketbra{\pm}{d}{\pm}{d}\hspace{1.6cm}\nonumber\\
+\frac{1}{128}(1-\eta_1)\left[1\mp2\textrm{Im}\left(\alpha\beta^*\right)\eta_2\right]\hspace{1.9cm}\nonumber\\
\times\ketbra{\mp}{a}{\mp}{a}\otimes\ketbra{\mp}{d}{\mp}{d}\hspace{1.5cm}\nonumber\\
+\frac{1}{64}\eta_1(1-\eta_2)\vert\alpha\vert^2\ket{0}^{(2)}_{ad}\langle0\vert^{(2)}_{ad}\hspace{2.45cm}\nonumber\\
+\frac{1}{64}\eta_1(1-\eta_2)\vert\beta\vert^2\ket{1}^{(2)}_{ad}\langle1\vert^{(2)}_{ad},\hspace{2.35cm}
\end{eqnarray}
where $\ket{\Psi_1'}_{ad}$ is the expected state as shown in
Eq.~(\ref{Z90:1}) and $\widetilde{\hat{\rho}}^{(+)}_{out}$
corresponds to the output states with no correction or with the
bit-flip correction, while $\widetilde{\hat{\rho}}^{(-)}_{out}$
corresponds to the output state after the phase-flip correction or
after both the phase-flip and the bit-flip correction.

The probability of success is given by
\begin{eqnarray}\label{Z90 Probability}
\widetilde{P}^{(\pm)}=\mbox{tr}\left(\widetilde{\hat{\rho}}^{(\pm)}_{out}\right)=\frac{1}{64}[1{\mp}(1-\eta_1)\eta_2\mbox{Im}(\alpha\beta^*)].
\end{eqnarray}

The fidelity is also given by
\begin{eqnarray}\label{Z90 fidelity}
\widetilde{F}^{(\pm)}=\frac{\langle\Psi'\vert_{ad}\widetilde{\hat{\rho}}^{(\pm)}_{out}\ket{\Psi'}_{ad}}{\mbox{tr}\left(\widetilde{\hat{\rho}}^{(\pm)}_{out}\right)},
\end{eqnarray}
where,
\begin{eqnarray}\label{Z90 fid}
\langle\Psi'\vert_{ad}\hat{\rho}^{(\pm)}_{out}\ket{\Psi}_{ad}=\hspace{5cm}\nonumber\\
\frac{1}{128}\left\{1+\eta_1-4\eta_1(1-\eta_2)\vert\alpha\beta\vert^2\right.\hspace{2cm}\nonumber\\
\left.{\mp}(1-\eta_1)\eta_2\mbox{Im}\left(\alpha\beta^*\right)\left[1{\mp}2\mbox{Im}\left(\alpha\beta^*\right)\right]\right\}.\
\end{eqnarray}
The average fidelity can also be given by
\begin{eqnarray} \label{Z90 average}
\widetilde{F}_{\textrm{ave}}&=&\frac{1}{4\pi}\int^{2\pi}_{0}d\phi\int^{\pi}_{0}\widetilde{F}^{(\pm)}\sin{\theta}d\theta\nonumber\\
&=&\frac{1}{4\pi}\int^{\frac{\pi}{2}}_{-\frac{\pi}{2}}d\phi\int^{\pi}_{0}\left(F^{(+)}+F^{(-)}\right)\sin{\theta}d\theta.
\end{eqnarray}

For a specific input state, the fidelity may be different in the
$Z_{90}$ operation compared with the re-encoder, but from
Eq.~(\ref{average}) and (\ref{Z90 average}) we can see that the
average fidelity is the same.

\section{Conclusions}
We have considered the problem of demonstrating the basic elements
of parity encoded linear optical quantum computation. We have shown
that operations on the smallest non-trivial example, the two-photon
parity state, can be demonstrated using a six-photon parametric
down-conversion source. Our proposal allows demonstration of basic
re-encoding, including explicitly the construction of the resource
state. The basic re-encoder is key in loss tolerant operation
\cite{ralph2005} as it acts as an error detector for loss. In effect
the re-encoder performs a quantum non-demolition measurement of
photon number on the parity qubit. The re-encoder can also be used
to implement arbitrary single qubit gates, which we also discussed
and the it can also be understood as a process of teleportation. We
have shown that multi-photon down-conversion in events can be
post-selected out of the data. We have also considered the effect of
mode-mismatch on the operation of our gates and shown that it leads
to an approximately linear reduction in fidelity as a function of
mode overlap.

Experimental demonstrations play a crucial role in evaluating and
testing the relative merits of different quantum processing schemes.
We hope that our proposal will stimulate such investigations of
parity state LOQC.

\acknowledgements YXG was funded by National Fundamental Research
Program (Grant No. 2006CB921907), National Natural Science
Foundation of China (Grant No. 60121503 and No. 60621064),
Innovation Funds from Chinese Academy of Sciences, International
Cooperate Program from CAS and Ministry of Science \& Technology of
China. AJFH and TCR were supported by the DTO-funded U.S.Army
Research Office Contract No. W911NF-05-0397 and the Australian
Research Council.

\bibliography{ref}
\end{document}